\documentclass[sigplan,10pt]{acmart}
\usepackage{microtype}

\PassOptionsToPackage{table, svgnames, x11names}{xcolor}
\usepackage[utf8]{inputenc} 
\usepackage[T1]{fontenc}    
\usepackage{hyperref}       
\usepackage{url}            
\usepackage{booktabs}       
\usepackage{amsfonts}       
\usepackage{nicefrac}       
\usepackage{microtype}      
\usepackage[linesnumbered,ruled,vlined]{algorithm2e}
\usepackage{fancyhdr} 
\usepackage{algpseudocode}
\usepackage{amsthm}
\usepackage{amsmath}
\usepackage{paralist}
\usepackage{setspace}
\usepackage{graphicx}
\usepackage{natbib}
\usepackage{subcaption}
\usepackage{multirow}
\usepackage{multicol}
\usepackage{makecell}
\usepackage{xspace}
\usepackage{svg}
\usepackage[table]{xcolor}
\usepackage{enumitem}
\usepackage[ruled,linesnumbered]{algorithm2e}
\usepackage{wrapfig}
\usepackage{float}
\usepackage{color,soul}
\usepackage{pbalance}
\usepackage{pifont}
\newcommand{\cmark}{\textcolor{green!55!black}{\ding{51}}}
\newcommand{\xmark}{\textcolor{red!70!black}{\ding{55}}}
\SetKwComment{Comment}{/* }{ */}

\usepackage[font=small,labelfont=small]{caption}

\newtheorem{defi}{Definition}[section]

\newcommand{\sysname}{JigSLO\xspace}

\renewcommand\footnotetextcopyrightpermission[1]{}

\title{\sysname: Solving the Resource Puzzle for Efficient Multi-LLM Serving}

\author{Kyungmin Bin$^{*}$, Seungbeom Choi$^{*}$, Daseul Bae$^{*}$, Jimyoung Son, Jieun Choi, Daehyeon Baek, \\Kihyo Moon, Minsung Jang, Hyojung Lee$^\dag$}
\affiliation{%
  \institution{Cloud Research Team, Samsung SDS}
   \country{} 
  }
\thanks{$^{*}$The first three authors contribute equally. $^\dag$Corresponding author (hyojung.lee.248@gmail.com).}

\renewcommand{\shortauthors}{Bin et al.}

\begin{abstract}

GPU spatial sharing co-locates instances with complementary demands, such as compute-bound prefill and memory-bound decode, promising cost-efficient multi-LLM serving.
Prior systems partition compute and coordinate shared memory capacity, but do not model bandwidth contention among co-located instances of heterogeneous LLMs.
This contention makes each instance's latency depend on the memory-access patterns of its co-runners, even when its own resource allocation remains unchanged.
Drawing on extensive measurements, we find that the latency impact of bandwidth contention can be captured by a time-shared memory-access model with contended and uncontended periods, so the overall latency can be analytically estimated as a time-weighted average of the latency estimates for these periods. 
We propose \sysname, a contention-aware multi-LLM serving system that manages memory resources and schedules each model's prefill and decode phases as separate instances, treating their co-location as a placement decision.
Its contention-aware predictor estimates per-instance latency under any co-location from standalone profiles alone.
Guided by these predictions, its placement algorithm jointly selects each instance's tensor parallelism, compute share, and batch configuration, pairing complementary memory-access patterns to maximize throughput within each model's latency targets.
\sysname's shared resource manager complements this on the capacity side, sharing weights, pooling KV cache, and transferring it between prefill and decode instances.
Across three datasets covering chat, code generation, and long-context summarization, \sysname outperforms state-of-the-art baselines by 2.3$\times$ in SLO attainment and 1.2$\times$ in throughput on average, matching their best attainment with half the GPUs on long-context workloads.
\end{abstract}

\begin{document}

\maketitle
\pagestyle{plain}
\section{Introduction}
\begin{figure}[t]
    \centering
    \includegraphics[width=\columnwidth]{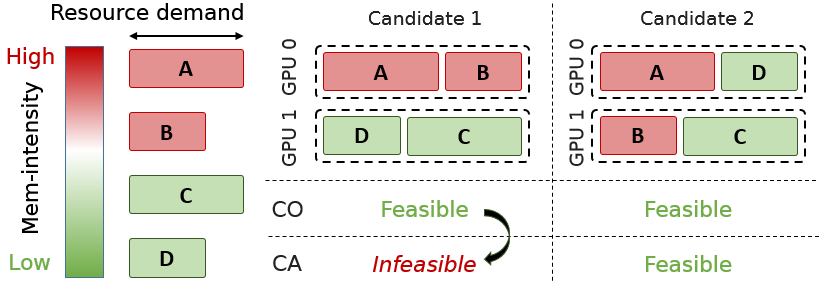}
    \caption{Two placement candidates for 4 instances on 2 GPUs. Instances differ in resource demands and memory-access intensity (left). Both candidates satisfy the compute and memory-capacity budgets, but candidate~1 packs the two memory-intensive instances (A, B) on the same GPU while candidate~2 separates them. A bandwidth contention-oblivious (CO) estimate deems both candidates feasible. Contention-aware (CA) prediction instead marks candidate~1 as infeasible because it is predicted to violate its latency SLO, and selects candidate~2.}
    \label{fig:concept}
\end{figure}

The demand for Large language model (LLM) is rapidly growing, driven by test-time scaling in reasoning models~\cite{openai2024o1, deepseek2025r1} and agentic workloads~\citep{claudecode, openai2025codex}, both of which generate far more tokens per query than prior applications. 
Sustaining high GPU utilization is essential to cost-efficient serving, but production deployments serve many LLMs concurrently~\cite{duan2024muxserve, xiang2025aegaeon, yu2025prism}, each as one or more serving instances whose resource demands vary with model size, precision, workload, and inference phase. A single instance rarely saturates a GPU, since a compute-bound instance leaves memory idle while a memory-bound one leaves compute idle. GPU spatial sharing, which co-locates instances with complementary resource demands on the same GPU, has emerged as a promising approach to cost-efficient multi-LLM serving.

Co-located instances share a GPU's compute and its memory resources. 
Compute resource can be partitioned, since NVIDIA MPS~\cite{nvidia_mps} bounds the fraction of streaming multiprocessors (SM) available to each instance.
Memory resource, however, is shared along two dimensions, capacity and bandwidth.
Capacity determines how much KV cache each instance can hold and hence its batch size, so co-located instances must be able to acquire and release it on demand as load shifts. 
Static partitions cannot follow such shifts, whether a per-instance memory limit that MPS fixes at launch or fixed static KV cache pool a serving engine reserves for each instance~\cite{tan2021serving, lee2024parvagpu}.
The bandwidth available to an instance, by contrast, is determined by its own memory-access intensity and that of its co-runners, bounded by the GPU's peak bandwidth. \textit{Contention} over this shared bandwidth thus makes an instance's latency depend on its co-runners even when its own compute share and memory capacity allocation remains unchanged. We observe that this contention may cause severe latency increase of serving instances under highly contended co-locations (\S\ref{sec:motivation}), making the instances violate latency SLOs.

Existing multi-LLM serving systems~\cite{duan2024muxserve, yu2025prism} place instances according to their compute or memory demand and leave this contention unmodeled. 
Latency prediction research for single-model serving does not fill the gap either, as it covers standalone execution~\cite{zhong2024distserve, agrawal2024vidur} or the contention between one model's prefill and decode~\cite{lin2026bullet, chen2026muxwise}, but not the contention among heterogeneous co-located instances. Placement in these systems thus cannot distinguish placement candidates that satisfy every compute and memory capacity constraint yet differ in the latency they actually deliver. 
In Figure~\ref{fig:concept}, for example, both candidates satisfy these constraints and appear SLO-feasible under bandwidth contention-oblivious (CO) placement. Candidate~1, however, co-locates the two memory bandwidth-intensive instances A and B, whereas candidate~2 separates them, and only a contention-aware (CA) placement reveals that candidate~1 would violate its SLO. Multi-LLM placement must therefore account for bandwidth contention alongside compute and capacity demands.

Contention-aware placement requires predicting each instance's latency under any co-location. Calibrating a contention model on co-located measurements is feasible for one model's prefill and decode~\cite{lin2026bullet} but not for the combinatorially many co-locations a heterogeneous multi-LLM placement must evaluate. We therefore aim to predict per-instance latency from standalone profiles alone (\S\ref{sec:latency_prediction}). Our extensive measurements show that this is possible because bandwidth contention reduces to time-shared memory access, so an instance's latency can be estimated as a time-weighted average of its latency in uncontended and contended periods (\S\ref{sec:motivation}). We predict uncontended latency using a model fitted to standalone profiles. For contended periods, we analytically estimate the bandwidth available to the instance under co-location and evaluate the same model at that bandwidth. We then take a weighted average of these latency estimates, with weights derived from co-runners' memory-access intensity, without requiring measurements of co-located execution.

Contention-aware placement must also account for GPU memory capacity, which constrains feasible batch sizes and instance configurations.
Each instance's KV cache capacity bounds its effective batch size, so co-located instances should
dynamically resize their KV caches as request rates and sequence lengths shift. \texttt{kvcached}~\cite{xing2025towards} provides such flexibility through elastic GPU memory allocation via CUDA Virtual Memory Management (VMM)~\cite{VMM}. In our evaluated configuration, it exhibits higher serving latency than our shared KV cache manager as the request rate increases (\S\ref{sec:eval_kvslab}). Our shared KV cache manager (\S\ref{sec:resource_manager}) instead preallocates a shared KV cache pool and coordinates block allocation through lightweight metadata updates, without modifying existing paged attention implementations.




Based on these insights, we present \sysname, a contention-aware multi-LLM serving system that schedules each model's prefill and decode phases as separate instances, predicts per-instance latency by modeling contention, and co-locates instances with complementary memory-access patterns. Our contributions are as follows.
\begin{compactenum}[$\circ$]
    \item \textit{A contention-aware latency predictor}
    analytically estimates per-instance latency under co-location by modeling bandwidth contention as time-shared memory access, without requiring exhaustive profiling of co-location combinations.
    \item \textit{An SLO-aware placement algorithm}
    maximizes per-GPU throughput under SLO constraints by querying the contention-aware predictor to jointly select each instance's tensor parallelism (TP) degree, SM allocation, batch configuration, and co-location.
    \item \textit{A shared resource manager}
    lets co-located instances draw GPU memory from a shared pool on demand, combining cross-instance weight sharing, a slab-based KV cache pool for heterogeneous block sizes, and prefill-to-decode KV cache transfer.
\end{compactenum}


We evaluate \sysname on two NVIDIA H100 nodes (16 GPUs total). Across three datasets, \sysname outperforms state-of-the-art baselines by 2.3$\times$ in SLO attainment and 1.2$\times$ in throughput on average, showing that contention-aware placement and shared resource management together make co-located multi-LLM serving cost-efficient.

\section{Background}\label{sec:background}

\noindent\textbf{LLM inference phases and instance.}
LLM inference proceeds in two phases with distinct resource characteristics. 
The prefill (P) phase processes prompt tokens in parallel and is often compute-bound for sufficiently large batched general matrix multiplication (GEMM) operations. Its primary latency metric is time-to-first-token (TTFT). The decode (D) phase generates tokens autoregressively and is often memory-bound at small batch sizes. Its primary latency metric is time-per-output-token (TPOT). Prior systems address the resource mismatch between the two phases either by interleaving prefill chunks with decode tokens via chunked-prefill scheduling~\citep{agrawal2024sarathi}, or by disaggregating them onto different GPUs~\citep{patel2024splitwise,zhong2024distserve} at the cost of KV cache transfer and underutilization when either phase is lightly loaded. \sysname instead deploys P/D separation by co-locating the two phases when their resource demands are complementary or disaggregating them otherwise.


We accordingly use \emph{instance} to denote a placement-level unit of LLM serving, characterized by the tuple (model, precision, TP degree, P/D phase). Each instance maintains one copy of model weights and may span multiple GPUs depending on its tensor parallelism degree. Determining whether a placement meets its SLOs therefore requires quantifying \textit{cross-instance contention}.

\smallskip
\noindent\textbf{GPU partitioning.}
NVIDIA GPUs provide two mechanisms for concurrent multi‑tenant execution. MIG~\citep{nvidia_mig} enforces hardware‑level partitions, assigning each slice a disjoint set of SMs, HBM capacity, bandwidth, and L2 cache, thereby offering strong isolation but only at a coarse granularity. 
In contrast, NVIDIA MPS~\citep{nvidia_mps} enables concurrent execution of multiple serving instances with limited isolation of compute resources. \sysname uses its active thread percentage setting to restrict the number of GPU threads each instance can use concurrently. These limits constrain compute usage without guaranteeing disjoint SM assignments. However, memory bandwidth remains shared without corresponding per-instance partitioning, creating the resource isolation asymmetry that motivates our contention analysis in~\S\ref{sec:motivation}.



\smallskip
\noindent\textbf{Dynamic KV cache management.}
Under MPS GPU partitioning, GPU memory capacity is shared across instances, while inter-process sharing and allocation of memory requires explicit coordination. Existing LLM serving frameworks (e.g., vLLM~\citep{kwon2023vllm}) statically partition memory across multiple instances by reserving a fixed pool per instance. This approach prevents the KV cache pool from dynamically adjusting to workload variation (e.g., changes in request rate or sequence length), capping the number of concurrent requests at runtime. Therefore, a flexible and dynamic KV cache management mechanism is essential to realize the full potential of GPU spatial sharing. We provide a detailed analysis of the two approaches (static and dynamic) in~\S\ref{sec:eval_kvslab}.

\section{Impact of Memory Bandwidth Contention on Multi-LLM Serving}\label{sec:motivation}
This section asks three questions about contention under GPU spatial sharing: its severity, its dependence on workload configuration, and its mechanism. The answers motivate \sysname's latency predictor and placement algorithm.

\begin{figure}[t]
  \centering
  \begin{subfigure}{0.49\linewidth}
    \includegraphics[width=\linewidth]{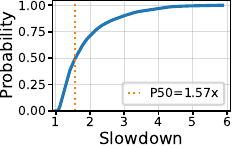}
    \caption{CDF of latency slowdown}
    \label{fig:motivation_slowdown_cdf}
  \end{subfigure}\hfill
  \begin{subfigure}{0.49\linewidth}
    \includegraphics[width=\linewidth]{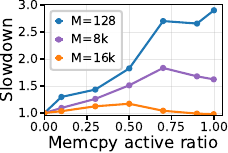}
    \caption{Slowdown of GEMMs}
    \label{fig:motivation_slowdown_GEMM}
  \end{subfigure}
  \caption{Contention severity and memory-subsystem pressure. (a) CDF of latency slowdown across the configuration sweep. 
  (b) Slowdown of a target GEMM kernel against interferer.
  }
  \label{fig:motivation}
\end{figure}

\smallskip
\noindent\textbf{Setup and terminology.} We profile co-located LLM instances on NVIDIA H100 SXM GPUs under MPS-based SM partitioning, using CUDA 13.0 and vLLM 0.13.0~\citep{kwon2023vllm}. We refer to the instance whose latency we measure as the \textit{target}, and any other instance sharing the GPU as an \textit{interferer}. Profile configurations cover three open models (Llama-3.1-8B/70B, Qwen3-32B), two precisions (BF16, FP8), both prefill and decode phases, input sequence lengths (ISL) up to 4096 and batch sizes (BS) up to 128, all sampled at powers of two, and TP degrees in \{1, 2, 4, 8\}. For each configuration tuple, we sweep SM allocations in 10\% increments, with co-located instances' allocation summing to 100\%. 
The sweep yields over 13{,}000 contended-latency samples for the analysis below. We define slowdown as the ratio of co-located latency to standalone latency under the same instance configuration and SM allocation.



\smallskip
\noindent\textbf{Observation 1. Contention is severe and highly variable.} Figure~\ref{fig:motivation_slowdown_cdf} shows the CDF of latency slowdown across the configuration sweep. The median slowdown is 1.57$\times$, while the tail reaches nearly 6$\times$ in the worst case. A naive co-location policy therefore risks severe SLO violations, motivating a contention-aware approach. 


\begin{figure}[t]
    \begin{subfigure}{0.48\columnwidth}
        \includegraphics[width=\columnwidth]{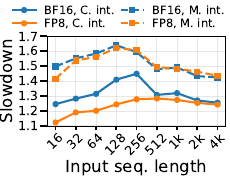}
        \caption{Slowdown of prefill targets}
        \label{fig:prefill_slowdown}
    \end{subfigure}
    \begin{subfigure}{0.48\columnwidth}
        \includegraphics[width=\columnwidth]{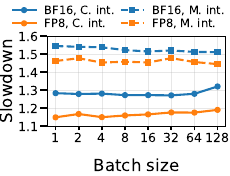}
        \caption{Slowdown of decode targets}
        \label{fig:decode_slowdown}
    \end{subfigure}
    \caption{
    Sensitivity of co-location tolerance to target configuration. For each target (prefill in (a), decode in (b)), we plot average slowdown over standalone execution at the same SM allocation, when co-located with a fixed compute-bound interferer (C. int.) (prefill of Llama-3.1-8B at ISL = 1024, BF16) or a fixed memory-bound interferer (M. int.) (decode of Llama-3.1-8B at BS = 2, BF16). SM allocations are averaged over 20--80\%.
    }
    \label{fig:motivation_workload}
    \end{figure}


\smallskip
\noindent\textbf{Observation 2. Slowdown is governed by a job's memory-boundedness, not its size alone.} 
Co-locating prefill (Figure~\ref{fig:prefill_slowdown}) and decode (Figure~\ref{fig:decode_slowdown}) targets with a fixed compute-bound or memory-bound interferer, 
we find that job size (the number of processed tokens) alone does not explain slowdown. The memory-boundedness of both the target and its interferer is also important.
The prefill target's slowdown peaks at an intermediate input length of 128 -- 256 and declines as the target becomes more compute-bound at longer sequences\footnote{Longer inputs amortize the cost of reading model weights over more tokens, raising arithmetic intensity and shifting prefill toward compute-bound execution.}. In contrast, the decode target's slowdown remains elevated. The interferer confirms this: the compute-bound interferer is the larger job, a 1024-token prefill against a batch-of-2 decode, yet being compute-bound it causes less slowdown than the small memory-bound interferer. FP8 lowers slowdown across the board by shrinking the memory footprint.

\begin{figure}[t]
    \centering
    \includegraphics[width=0.9\columnwidth]{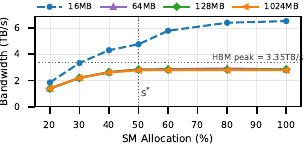}
    \caption{HBM bandwidth achieved (bytes read $/$ time) by the streaming-copy kernel across SM allocations (20 -- 100\,\%) under different buffer footprints (16 -- 1024\,MB). Bandwidth scales with SM share up to roughly $50\,\%$, then saturates at  $87\,\%$ of the $3.35$\,TB/s HBM peak bandwidth. We denote the saturation point as $\text{s}{\text{*}}$.}
    \label{fig:sm_gating}
\end{figure}

\smallskip
\noindent \textbf{Observation 3. HBM bandwidth scales with SM share.}
MPS partitions SMs across processes but not the memory subsystem, so the amount of HBM
bandwidth a process obtains at a given SM allocation is not straightforward.
To characterize this, we use a streaming-copy kernel with four different buffer footprints (16\,MB, 64\,MB, 128\,MB, and 1024\,MB). 
Except for the 16\,MB, the footprints exceed the H100's 50\,MB L2 capacity and cannot fit entirely in L2.
We run the streaming-copy kernel alone at full duty cycle across SM allocations from 20\,\% to 100\,\%. Figure~\ref{fig:sm_gating} shows that (i) for footprints exceeding L2 capacity, HBM bandwidth scales linearly with SM share up to roughly 50\,\% (denoted by $\text{s}{\text{*}}$), then saturates at $\sim87\,\%$ of the $3.35$\,TB/s peak, and (ii) for the footprint smaller than L2 capacity, memory bandwidth behaves similarly to the other cases but also exceeds the HBM peak since the L2 cache is engaged~\cite{elvinger2025understanding}. In the sub-saturated regime ($\text{SM allocation} < \text{s}{\text{*}}$), the streaming-copy kernel's achieved HBM bandwidth is approximately proportional to SM allocation. This relationship provides the standalone bandwidth basis for \sysname's contention-aware latency model (\S\ref{sec:latency_prediction}). 



\smallskip
\noindent\textbf{Observation 4. Slowdown grows approximately linearly with interferer duty cycle.} To isolate how an interferer's memory traffic affects a target kernel's latency, we co-locate a target GEMM kernel (an MLP projection of Llama-3.1-8B with hidden 4096 and intermediate 14336, BF16) with a token-dimension size $M$. We run the target alongside a streaming-copy interferer and sweep its memcpy ratio, defined as the fraction of time the interferer actively generates memory traffic throttling the amount of bandwidth left for the target (Figure~\ref{fig:motivation_slowdown_GEMM}). With SMs split 50/50, compute-bound targets exhibit little slowdown regardless of the interferer's memcpy active ratio, whereas memory-bound targets show approximately linear increases in slowdown as this ratio rises. The slope therefore depends on the target's memory-boundedness, with slowdown reaching 1.01$\times$ for $M=16$k (compute-bound) and 2.9$\times$ for $M=128$ (memory-bound).
These results indicate that contention depends on both the target's sensitivity to available memory bandwidth and the interferer's memory-access activity. This linear trend supports approximating co-located execution as periods with and without bandwidth contention, motivating a time-shared memory-access model that combines the corresponding latency estimates according to the fraction of time spent in each period.



\smallskip
\noindent\textbf{Implications.} 
Memory bandwidth contention varies substantially across instance configurations, so placement decisions must account for how co-located instances affect one another's latency.
The observations identify three relevant factors: the target's sensitivity to reduced bandwidth, the interferers' memory-access activity, and the SM allocation of each instance. The measured relationships between SM allocation and achievable bandwidth, and between interferer duty cycle and slowdown, support a time-shared memory-access model. This motivates \sysname's contention-aware latency predictor~(\S\ref{sec:latency_prediction}).

\section{\sysname System Design}\label{sec:system}

\begin{figure}[t]
  \centering
  \includegraphics[width=\linewidth]{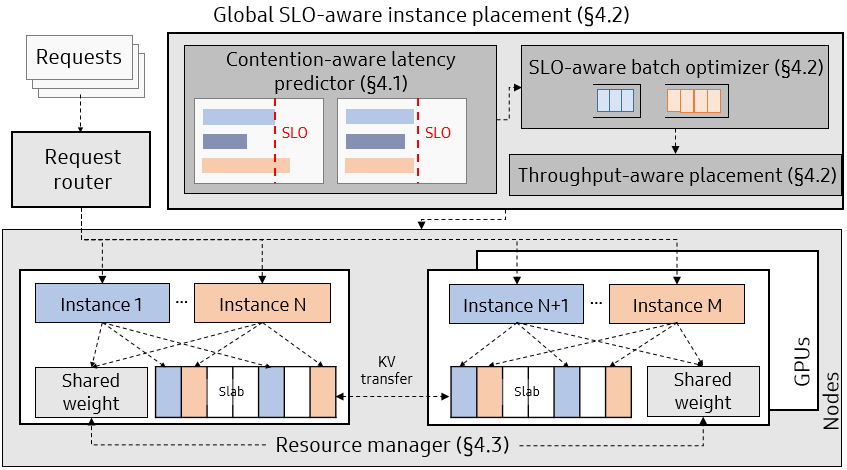}
  \caption{\text{\sysname architecture.}}
  \label{fig:overview}
\end{figure}

Figure~\ref{fig:overview} shows the overall architecture of \sysname. Given a set of LLMs with workload specifications and SLO targets, \sysname operates in two stages. An offline placement stage uses the contention-aware latency predictor~(\S\ref{sec:latency_prediction}) to drive the SLO-aware instance placement algorithm~(\S\ref{sec:placement}), determining how to co-locate instances across GPUs. An online serving stage then dispatches requests according to the placement, while the shared resource manager~(\S\ref{sec:resource_manager}) coordinates KV cache and weight sharing across each GPU.

\subsection{Contention-aware Latency Prediction}\label{sec:latency_prediction}
The predictor takes a co-location, a set of instances sharing a GPU, and returns each instance's per-step latency under it, the time of one forward pass of the instance over its batch. 
Each instance is specified by its model configuration (model, precision, phase, TP degree, SM allocation) and batch configuration (batch size, input and output sequence lengths). 
For a decode instance, the latency is averaged over the output sequence length, since each step reads a growing KV cache. 
The predictor is analytical, so evaluating a co-location is inexpensive and requires no measurement beyond standalone profiles, which lets the placement algorithm~(\S\ref{sec:placement}) explore many co-locations and batch sizes. We first model the latency of an instance running alone, then derive its latency under co-location from that model. 


\smallskip
\noindent \textbf{Standalone latency.}
The standalone latency of an instance $t$ depends on its model and batch configuration. 
We model it as a linear combination of four kernel-level latency estimate components, namely memory-bound GEMMs, compute-bound GEMMs, attention, and TP communication, which scale differently with SM allocation:
\begin{equation*}
{L}^{\text{std}}(t) = \alpha \cdot {L}^{\text{mem}}_{\text{gemm}}(t) + \beta \cdot {L}^{\text{comp}}_{\text{gemm}}(t) + \gamma \cdot {L}_{\text{attn}}(t) + \delta \cdot {L}_{\text{tp}}(t) + \epsilon.
\end{equation*}
The three kernel features ($L_{\text{gemm}}^{\text{mem}}$, $L_{\text{gemm}}^{\text{comp}}$, $L_{\text{attn}}$) are roofline-based estimates obtained by applying LLMViewer's per-layer analysis~\cite{yuan2024llm} to the instance's model, precision, phase, batch size, and sequence lengths, 
with the TP degree, compute capacity scaled by SM share, and bandwidth scaled by the attainable fraction of Observation~3. Memory-bound GEMMs and attention depend on memory bandwidth, while compute-bound GEMMs primarily depend on compute capacity (Observations~2 and~3), so we model them separately. The TP feature ($L_{\text{tp}}$) is the NVLink transfer time of the activations exchanged by tensor-parallel collectives. The intercept $\epsilon$ absorbs configuration-independent overheads such as kernel launches.
Because these features are theoretical bounds, we fit one set of coefficients per (model, precision, phase) to standalone latencies measured across SM allocations, batch sizes, and sequence lengths~(\S\ref{sec:motivation}).

\smallskip
\noindent \textbf{Contention-aware latency prediction.}
Observation 4 showed that slowdown grows approximately linearly with interferer duty cycle, motivating a time-shared memory-access model of bandwidth contention.
Let $\mathcal{C}$ be the instances sharing a GPU and $t \in \mathcal{C}$ the target. We approximate each interferer's memory-access activity as an on-off process, with active and inactive periods of bandwidth consumption. The subset of active interferers, $\mathcal{S} \subseteq \mathcal{C}\setminus\{t\}$, defines a \textit{contention state}, and the target's available bandwidth depends on this subset.
As the contention state varies over time, we approximate the target's expected latency by the time average of its per-state latency estimates. 

\begin{defi}[Contended latency]\label{def:contended}
The contended latency of target instance $t$ is defined by: 
    \begin{equation*}
    {L}^{\text{cont}}(t) = \sum_{\mathcal{S} \subseteq \mathcal{C} \setminus \{t\}} \mathbf{P}(\mathcal{S}) \cdot {L}^{\text{cont}}(t \mid \mathcal{S}),
\end{equation*}
where $\mathbf{P}(\mathcal{S})$ is the fraction of time in which $\mathcal{S}$ is the active set and ${L}^{\text{cont}}(t \mid \mathcal{S})$ is the target's estimated latency under the bandwidth available in that state.
\end{defi}
\noindent Both terms are computed from ${L}^{\text{std}}$ and analytical memory-traffic estimates, without fitting to measurements of co-located execution.
We derive the two terms in turn.

\smallskip
\noindent \textbf{Deriving the two terms.}
The first term, $\mathbf{P}(\mathcal{S})$, follows from the interferers' memory-access activity. Each instance $i \in \mathcal{C}$ is an on-off consumer of bandwidth with active-time fraction $u_i = V_i / (L^{\text{std}}(i) \cdot B)$, the share of its step spent moving its theoretical memory traffic $V_i$ (as in LLMViewer~\cite{yuan2024llm}) at the device's peak bandwidth $B$. Treating interferers as independent, $\mathbf{P}(\mathcal{S})$ is the probability that those in $\mathcal{S}$ are active and all remaining interferers are inactive:
\begin{equation*}
\mathbf{P}(\mathcal{S}) = \prod_{i \in \mathcal{S}} u_i \cdot \prod_{i \in \mathcal{C}\setminus(\{t\}\cup \mathcal{S})} (1- u_i).
\end{equation*}

The second term, ${L}^{\text{cont}}(t \mid \mathcal{S})$, follows from the bandwidth the target retains in state $\mathcal{S}$. Let $p_i = \min(s_i / s\text{*}, 1)$ be the fraction of peak bandwidth attainable by instance $i$ in standalone execution at its SM allocation $s_i$, with the knee $s\text{*}$ profiled offline from the SM-allocation curve (Figure~\ref{fig:sm_gating}).

The target's share under contention state $\mathcal{S}$ is then
\begin{equation*}
p_{t \mid \mathcal{S}} =
\begin{cases}
p_t & \text{if } \sum_{i \in \{t\} \cup\mathcal{S}} p_i \leq 1,\\[2pt]
\text{s}_t / \textstyle\sum_{i\in \{t\} \cup\mathcal{S}} \text{s}_i & \text{otherwise.}
\end{cases}
\end{equation*}
When the standalone bandwidth shares of the target and active interferers sum to at most 1, the target retains its standalone bandwidth share. Otherwise, we approximate bandwidth allocation as proportional to SM allocations of the target and active interferers. We then evaluate the standalone model using the bandwidth available in state $\mathcal{S}$, $L^{\text{cont}}(t \mid \mathcal{S}) = L^{\text{std}}(t, p_{t \mid \mathcal{S}}/p_t)$, where the second argument specifies the available bandwidth relative to standalone execution. We recompute the kernel-level regression features under this bandwidth and apply the same coefficients fitted to standalone measurements. In the uncontended state, $\mathcal{S}=\emptyset$, this reduces to $L^{\text{std}}(t)$. 
We evaluate the accuracy of both standalone and contended latency against measured latencies in~\S\ref{app:prediction_results}.

\smallskip
\noindent \textbf{Use in placement.} The placement algorithm~(\S\ref{sec:placement}) queries the predictor for each candidate co-location and batch configuration it examines, and converts the returned per-step latencies into TTFT and TPOT estimates that it checks against each model's SLO.

\subsection{SLO-aware Instance Placement}\label{sec:placement}

\begin{algorithm}[t]
  \caption{SLO-aware GPU Placement}
  \label{al:gpu_placement}
  \small
  \DontPrintSemicolon
  \LinesNumbered

  \KwIn{LLM list $N$, SLO constraints $O$, Num. of GPUs $G$}
  \KwOut{Placement $\mathcal{P}$}

  \Comment{Step 1: candidate generation}
  $C \leftarrow$ \textsc{EnumerateCombinations}($N$)\;
  \ForEach{$c \in C$}{
    $c.\text{config} \leftarrow$ \textsc{BatchOptimizer}($c$, $O$, $G$)\;
    \lIf{$c.\text{config} = \emptyset$}{remove $c$ from $C$}
  }

  \Comment{Step 2: candidate selection}
  $r \leftarrow G$;\quad $\mathcal{P} \leftarrow \emptyset$;\quad $\mathcal{A} \leftarrow \emptyset$\;
  \While{$\exists\, c \in C$ with $c.\text{gpus} \leq r$}{
    $c^{*} \leftarrow \arg\max$ \textsc{Priority}($c$, $\mathcal{P}$, $\mathcal{A}$) over such $c$\;
    $r \leftarrow r - c^{*}.\text{gpus}$;\quad $\mathcal{P} \leftarrow \mathcal{P} \cup \{c^{*}\}$\;
    $\mathcal{A} \leftarrow$ \textsc{AttainableThroughput}($\mathcal{P}$)\;
  }
  assign servers and GPUs to $\mathcal{P}$\;
  \Return $\mathcal{P}$

  \BlankLine
  \SetKwProg{Fn}{Procedure}{:}{}
  \Fn{\textsc{BatchOptimizer}($c$, $O$, $G$)}{
    $[low_i, high_i] \leftarrow$ \textsc{FeasibleRange}($c$, $i$, $O$), $\forall i \in c$\;

    
    $\mathbf{b} \leftarrow [low_i]$\ $\forall i \in c$\;
    \While{$\exists\, i$ with $low_i \leq high_i$}{
      $i \leftarrow$ lowest-throughput such instance; \,
      $m \leftarrow \lfloor (low_i{+}high_i)/2 \rfloor$\;
      $\mathbf{b}' \leftarrow \mathbf{b}$,
      $b'_i \leftarrow m$\;
      \lIf{\textsc{SatisfiesSLO}($c$, $\mathbf{b}'$, $O$) $\wedge$ \textsc{FitsInMemory}($c$, $\mathbf{b}'$, $G$)\;
      }
          {$\mathbf{b} \leftarrow \mathbf{b}'$; $low_i \leftarrow m+1$}
      \lElse{$high_i \leftarrow m{-}1$}
    }
    \lIf{\textsc{SatisfiesSLO}($c$, $\mathbf{b}$, $O$)
     $\wedge$ \textsc{FitsInMemory}($c$, $\mathbf{b}$, $G$)}
    {\Return $\mathbf{b}$}
\Return $\emptyset$\;
  }

\end{algorithm}
Guided by the latency predictor~(\S\ref{sec:latency_prediction}), \sysname's placement algorithm searches for instance configurations and co-locations that maximize per-GPU throughput while meeting each model's SLO and sustaining its request rate, treating bandwidth contention as a constraint alongside compute and memory capacity.

The unit of placement is the \textit{instance}, a tuple of (model, phase, TP degree, SM allocation, batch configuration), and each (model, phase) may have several instances to scale throughput via data parallelism. We assume each model's workload pattern (request rate, input/output sequence length distribution) is provided as input. 
Choosing these tuples for each \textit{instance}, deciding which instances to co-locate, and assigning them to GPUs are coupled and combinatorially hard to solve jointly, because a co-location fixes both the memory left for its peers and the contention-induced latency that decides SLO satisfaction. We therefore use the two-step heuristic of Algorithm~\ref{al:gpu_placement}, built around the \textit{candidate}, a set of co-located instances that share the same GPUs and hence the same TP degree, treated as the atomic placement unit. This granularity shrinks the search space while preserving the flexibility to find combinations that meet our objective. \textit{Candidate generation} enumerates candidates and selects a feasible batch configuration for each, discarding those with none. \textit{Candidate selection} then greedily picks the subset that maximizes per-GPU throughput for a given number of GPUs. 

\smallskip
\noindent \textbf{Step 1: Candidate generation.}
\textsc{EnumerateCombinations} (line~1) builds the candidate set $C$ by enumerating model, phase (prefill, decode), TP degree (1, 2, 4, 8), and SM allocation (20--100\%) over up to four instances per candidate, leaving the batch configuration ($c.\text{config}$) unspecified. 
Each instance's batch configuration is specified by a \textit{token budget} $b_i$, the total number of tokens processed per step across concurrent requests. 
For decode, this equals the batch size because each request contributes one token per step.
For prefill, it specifies the aggregate number of prompt tokens processed in a chunked prefill step. 

\textsc{BatchOptimizer} (line 3) searches for feasible token budgets for the instances in a candidate $c$, denoted as $\mathbf{b}$, aiming to improve per-GPU throughput under SLO and memory-capacity constraints. It first initializes each instance's search range $[low_i, high_i]$ and each token budget as $low_i$ using \textsc{FeasibleRange} (line~13--14). The range depends on the instance's phase. For decode, it spans 1 to the largest batch size profiled. For prefill, it covers token budgets whose TTFT, estimated using the latency predictor in \S\ref{sec:latency_prediction} as per-step latency multiplied by step count, satisfies the SLO.

The procedure searches these ranges using binary search (lines~15--20), prioritizing the instance with the lowest throughput among those with an unresolved range. At each iteration, it tests the midpoint of the selected instance's range while retaining the other instances' current token budgets. \textsc{SatisfiesSLO} checks whether all instances meet their phase-specific SLOs using the predicted latencies under the proposed configuration, while \textsc{FitsInMemory} checks the candidate's memory requirements (line~18). If both checks pass, the configuration is updated and the search proceeds toward larger budgets. Otherwise, it proceeds toward smaller budgets. This search is motivated by non-decreasing per-instance throughput until compute saturation and by feasibility only tightening as the token budget grows within the specified ranges. If no feasible budget configuration is found, \textsc{BatchOptimizer} returns $\emptyset$, and the candidate is removed from $C$ (line~4).

\smallskip
\noindent \textbf{Step 2: Candidate selection.}
Rather than exhaustively searching combinations of candidates, \sysname constructs the placement $\mathcal{P}$ greedily, adding one candidate at a time within the available $G$ GPUs. To guide each selection toward improving overall serving throughput, \textsc{Priority} ranks candidates by the additional throughput they contribute to the current placement per GPU required (line 7). At each iteration, we select the highest-priority candidate that fits within the remaining $r$ GPUs and update the attainable throughput $\mathcal{A}$ for each model phase (line 8--9). This process continues until no candidate fits. The selected candidates are then assigned to servers with enough available GPUs (line 10).

\textsc{Priority} ranks candidates by the increase in \textit{effective model throughput}, summed across models and divided by the number of GPUs the candidate requires. For each model, we sum the throughputs of its instances separately for prefill and decode, measured in requests per second. We cap each phase's total at the model's request rate because additional capacity beyond that rate does not serve more requests. We call this capped value the \textit{effective phase throughput}. Since every request requires both phases, the smaller of the two values determines the model's effective model throughput.
Because this throughput remains zero for a model until both phases are placed, we break ties by first favoring candidates that supply a missing phase, then comparing the increase in effective phase throughput per additional GPU.

\subsection{Shared Resource Manager}\label{sec:resource_manager}
To support the selected placement during serving, the shared resource manager coordinates memory allocation and sharing across the co-located instances. Naive memory allocation across co-located LLM instances wastes memory capacity in three ways, which \sysname's resource manager addresses in turn: (i) weight duplication eliminated via weight sharing, (ii) external fragmentation from heterogeneous KV block sizes resolved by a slab-based shared KV cache pool, and (iii) redundant KV blocks between P/D instances of the same model avoided via KV reuse or transfer depending on whether they share a GPU.




\smallskip
\noindent\textbf{Weight sharing.}
When the placement algorithm~(\S\ref{sec:placement}) co-locates multiple instances of the same model on a GPU, independent weight loading duplicates weight storage and reduces the memory available for KV cache. \sysname instead shares weight tensors across same-model instances via CUDA IPC handles. For each model on each GPU, the resource manager retains CUDA IPC handles for the weight tensors loaded by the first instance. Subsequent instances of the same model use these handles to map the existing allocations into their address spaces.


\begin{figure}
    \centering
    \includegraphics[width=1.0\columnwidth]{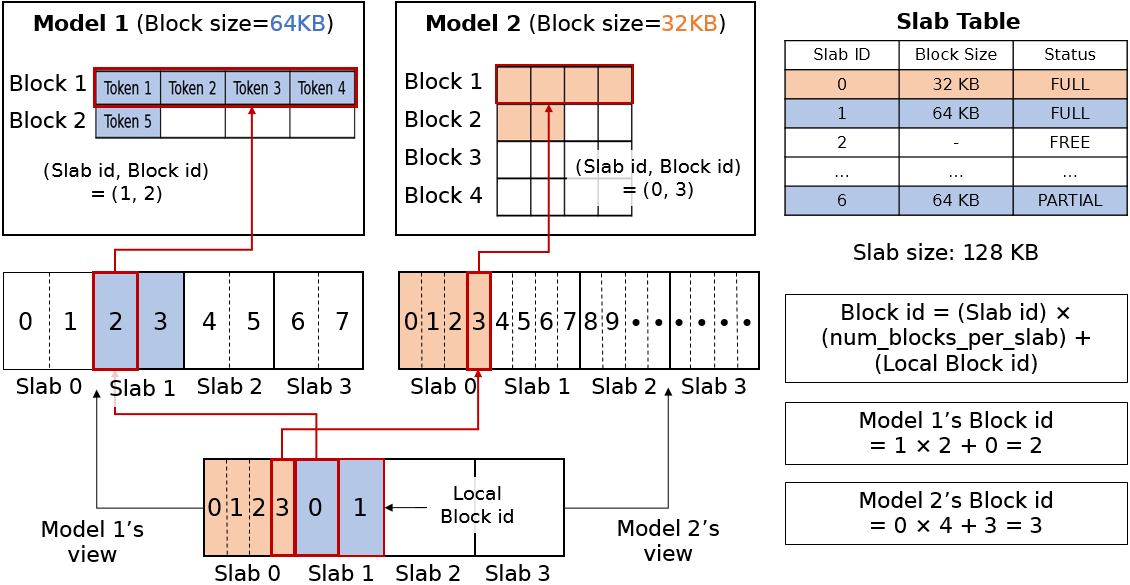}
    \caption{The \textit{KV slab manager} enables two co-located models with heterogeneous block sizes (Model 1: 64\,KB, Model 2: 32\,KB) to share a single physical KV tensor managed as fixed-size 128\,KB slabs.
    The slab table tracks each slab's block size and status. 
    At runtime, each instance sees block IDs composed from (slab ID, local block ID), allowing the same physical memory to back blocks of different sizes for different models.
    }
    \label{fig:kvslab}
\end{figure}


\smallskip
\noindent\textbf{Shared KV cache coordination (KV slab manager).}
\sysname reserves GPU memory for weights, peak activations, and other runtime needs, and uses the remaining capacity as a shared KV cache pool. Co-located models can have different KV block sizes due to their attention architecture (number of KV heads, head dimension) and KV cache precision, and frequently allocating and freeing such variable-sized blocks risks external fragmentation, calling for a fragmentation-safe pooling strategy.

One approach is on-demand allocation, as in \texttt{kvcached}. \sysname instead reserves the pool once and coordinates allocation purely through block-ID mapping, keeping the critical path free of driver overhead. To avoid fragmentation under variable block sizes, the pool is managed by a slab allocator of fixed-size chunks, each re-formattable to any co-located model's KV block size at runtime.

\textit{The KV slab manager} (\texttt{kvslab}) coordinates the shared physical KV tensor by composing each instance's block IDs from (slab ID, local block ID) pairs (Figure~\ref{fig:kvslab}). 
All slabs on a GPU have the same byte size, set to a multiple of every co-located model's KV block size so any slab can be formatted to any of those sizes. A slab table records each slab's block size and allocation status (\texttt{FREE}, \texttt{PARTIAL}, \texttt{FULL}).
%
On allocation, the manager finds a slab of matching block size or carves a new one from the pool, and on release, empty slabs return to the pool, so its composition tracks the co-located workload mix without statically partitioning the pool among instances.

\smallskip
\noindent\textbf{KV cache transfer between P/D instances.}
When phase separation places prefill and decode instances of the same model, the prefill output's KV must reach the decode instance for autoregressive generation. We handle this differently depending on whether the two instances share a GPU.

For \textit{intra-GPU} transfers, \sysname reuses the underlying physical blocks rather than copying. Since prefill and decode instances of the same model share a unified view of the underlying physical tensor through the slab manager, passing the KV block IDs and associated metadata to the decode instance suffices. The KV cache transfer cost thereby reduces to a metadata update with negligible overhead.

For \textit{inter-GPU} transfers, \sysname stages KV blocks through Mooncake Store~\citep{qin2025mooncake}, a host-resident KV cache store that exposes blocks across GPUs via content keys. Prefill writes its blocks to the host store under a per-request key, the prompt router assigns a decode instance, and decode pulls the blocks via RDMA.
\section{Implementation}\label{sec:implementation}
We implement \sysname in approximately 13K lines of Python on top of vLLM 0.13.0~\cite{kwon2023vllm}. A frontend router forwards requests to serving instances via shared memory to minimize inter-process communication overhead.
The contention-aware predictor and the placement algorithm run offline as standalone Python modules. A custom harness collects the standalone profiles of~\S\ref{sec:latency_prediction} that the predictor is fit to. The placement module's output is a configuration file from which the launcher spawns serving instances, enforcing each instance's SM allocation by setting \texttt{CUDA\_MPS\_ACTIVE\_THREAD\_PERCENTAGE} per process.

The shared resource manager runs as a per-node process that serves instances over ZeroMQ~\cite{ZMQ} and plugs into vLLM through three of its extension interfaces. The weight-share manager provides a custom model loader registered through vLLM's loader registry, mapping already-resident weight tensors into later same-model instances. The KV slab manager replaces vLLM's \texttt{BlockPool} with a \texttt{SlabBlockPool} of the same interface: it allocates a single shared KV tensor at initialization, exposes it to every co-located instance via CUDA IPC, and hands out and reclaims block IDs at runtime. KV cache movement is a custom KV connector, which is a no-op for an intra-GPU P/D pair---the slab manager already gives both a unified block-ID space---and moves blocks over RDMA via Mooncake Store~\cite{qin2025mooncake} otherwise.

\section{Evaluation}\label{sec:evaluation}

\subsection{Experimental Setup}\label{sec:setup}

\textbf{Environment.} All experiments are conducted on a two-node cluster (each equipped with 8 GPUs, Table~\ref{tab:testbed}). Each node runs Ubuntu 22.04 with NVIDIA driver 580 and CUDA 13.0. We evaluate three open-source LLMs spanning 8B-70B parameters (Llama-3.1-8B, Qwen3-32B, Llama-3.1-70B), each in BF16 and FP8 precision. The KV cache experiments in~\S\ref{sec:eval_kvslab} additionally use Qwen3-8B. The number of models served per variant differs across experiments. 

\smallskip
\noindent\textbf{Datasets.} We use three datasets covering distinct workload patterns: Azure-Code~\cite{patel2024splitwise} (production code generation traces), ShareGPT~\citep{sharegpt2023} (chat) and ArXiv Summarization~\citep{cohan2018arxivsum} (long-context summarization). Request arrivals follow a Poisson process with configurable rate. 

\begin{table}
  \centering
  \small
  \begin{tabular}{ll}
    \toprule
    Component & Specification \\
    \midrule
    GPU & 8$\times$ H100 SXM5 80 GB \\
    NVLink & 4.0 / NVSwitch \\
    NICs & Mellanox ConnectX (RDMA) \\
    CPU & 2$\times$ Xeon 8468 (96 cores) \\
    DRAM & 2 TB DDR5-4800 \\
    \bottomrule
  \end{tabular}
  \caption{Per-node testbed configuration. The two-node cluster has 16 H100 GPUs in total, interconnected by 400 Gbps NDR InfiniBand.}
  \label{tab:testbed}
  \end{table}

\begin{table}
  \centering
  \small
  \begin{tabular}{lccc}
    \toprule
    & Azure-Code & ShareGPT & ArXiv Sum. \\
    \midrule
    P90 input length & 5187 & 405 & 16153 \\
    P90 output length & 55 & 665 & 420 \\
    TTFT/token & 1.5\,ms & 3.0\,ms & 1.0\,ms \\
    TTFT SLO & 7.8\,s & 1.2\,s & 16\,s \\
    \bottomrule
  \end{tabular}
  \caption{Per-dataset SLO targets. TTFT SLO is derived from P90 input length $\times$ TTFT/token.
  TPOT SLO is 50 ms.
  }
  \label{tab:slo_targets}
  \end{table}

\begin{figure*}[t]
    \centering
    \includegraphics[width=0.95\textwidth]{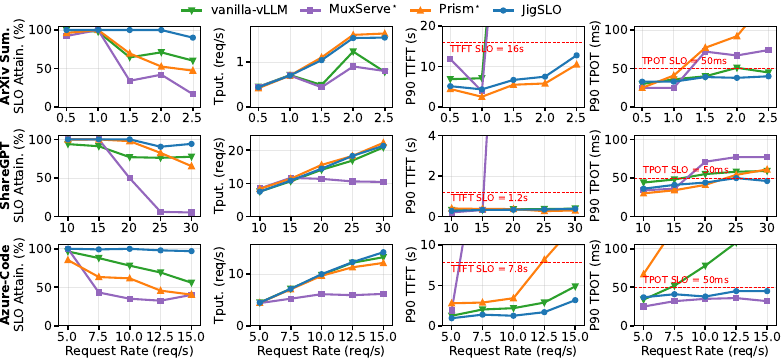}
    \caption{End‑to‑end performance of all baselines across three datasets. Columns represent (from left to right): SLO attainment rate (\%), request throughput (req/s), P90 TTFT (s), and P90 TPOT (ms) tail latency.}
    \label{fig:e2e_main}
\end{figure*}

\smallskip
\noindent\textbf{Metrics and SLOs.} We report four metrics: P90 TTFT and P90 TPOT as tail‑latency measurements, SLO attainment, and request throughput. SLO attainment is the fraction of requests that satisfy both the TTFT and TPOT SLOs, while request throughput is the number of completed requests per second. All metrics are averaged over the served models. We enforce a uniform TPOT SLO of 50 ms for every dataset and employ dataset‑specific TTFT SLOs that follow the convention commonly adopted in prior works~\cite{xiang2025aegaeon,lin2026bullet}. For each dataset, the TTFT SLO is calculated as the 90‑percentile token length multiplied by a per‑token processing time. Detailed SLO targets for each dataset are listed in Table~\ref{tab:slo_targets}.



\smallskip
\noindent\textbf{Baselines.} We compare \sysname against three baselines including state-of-the-art multi-LLM serving systems in order to assess the end-to-end performance (Table~\ref{tab:baselines}).

\begin{table}[h]
\centering
\small
\setlength{\tabcolsep}{4pt}
\begin{tabular}{lccl}
\toprule
\textbf{System} & \textbf{Placement} & \textbf{P/D} & \textbf{KV cache} \\
\midrule
vanilla-vLLM~\cite{kwon2023vllm}        & contention-oblivious    & \cmark & static \\
MuxServe$^\star$~\cite{duan2024muxserve} & greedy (compute) & \cmark & \texttt{kvslab} \\
Prism$^\star$~\cite{yu2025prism}         & greedy (memory)      & \xmark & \texttt{kvcached} \\
\sysname                                  & contention-aware        & \cmark & \texttt{kvslab} \\
\bottomrule
\end{tabular}
\caption{Summary of each baseline. Each column (starting from left to right) denotes the placement strategy, whether prefill/decode disaggregation is applied, and the KV cache management scheme.}
\label{tab:baselines}
\end{table}


\begin{compactenum}[$\circ$]
    \item Vanilla-vLLM~\cite{kwon2023vllm} represents the direct extension of a single-model serving engine to the multi-LLM setting. We statically partition the KV cache across instances to mirror vLLM's default behavior. 
    For placement, this baseline uses standalone latency estimates in \sysname's placement algorithm, a \textit{contention-oblivious} (CO) placement, while retaining prefill/decode disaggregation. The effects of contention-aware prediction and shared KV cache management are further explored as ablation experiments in ~\S\ref{subsec:ablation}.
    We enable cross-instance weight sharing, without which placement would be prohibitively constrained on multi-instance same-model deployments.
    \item MuxServe$^\star$~\cite{duan2024muxserve} is a multi-LLM serving system whose global placement runs an enumeration-based greedy algorithm that co-locates LLMs on GPU meshes, prioritizing computation heavy models, with the runtime giving priority to prefill over decode on the same mesh.
    \item Prism$^\star$~\cite{yu2025prism} runs a greedy placement that balances per-GPU KV memory pressure relative to SLO headroom, using the \texttt{kvcached} library~\cite{xing2025towards} for elastic KV memory and a local scheduler that coordinates requests among co-located models via priority-based admission control.
\end{compactenum}

We re-implement both MuxServe and Prism on top of \sysname's execution engine and resource manager (denoted by the $\star$ suffix), to ensure a fair comparison by eliminating performance differences caused by disparate engines: MuxServe targets an outdated vLLM version (0.2.0) and Prism is built on SGLang~\cite{zheng2023sglang}.

\begin{figure*}[t]
    \centering
    \begin{subfigure}{0.49\textwidth}
        \includegraphics[width=\textwidth]{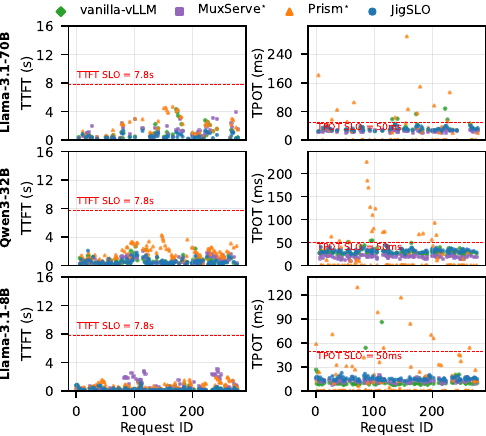}
        \caption{Azure dataset at request rate 5}
    \end{subfigure}
    \begin{subfigure}{0.49\textwidth}
        \includegraphics[width=\textwidth]{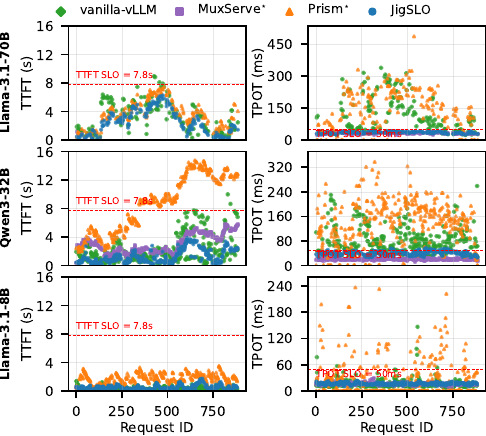}
        \caption{Azure dataset at request rate 15}
    \end{subfigure}
    \caption{Per-request TTFT and TPOT for three models on Azure-Code at aggregate request rates of 5\,req/s (a) and 15\,req/s (b). At 15\,req/s, the baselines exhibit TTFT or TPOT SLO violations, while \sysname satisfies both SLOs. In (b), MuxServe$^\star$'s TTFT points for Llama-3.1-8B and Llama-3.1-70B are omitted because all exceed the plotted limit of 16\,s.}
    \label{fig:req_stats}
\end{figure*}

\subsection{End-to-end Performance}\label{subsec:end-to-end_performance}
We evaluate \sysname's end-to-end performance through three experiments measuring SLO attainment, throughput, and tail latency: (i) sweeping the offered request rate at fixed cluster size, (ii) scaling cluster size at fixed total request rate, and (iii) scaling the number of models at fixed request rate per model. Unless otherwise specified, request rate refers to the aggregate arrival rate across all evaluated models.

\smallskip
\noindent\textbf{Request rate scaling.}
We deploy FP8-quantized variants of Llama-3.1-8B, Qwen3-32B, and Llama-3.1-70B concurrently across four GPUs. Each model is served by multiple co-located instances rather than a single one. Llama-3.1-70B receives half the request rate of each of the other two models. For example, an aggregate request rate of 10\,req/s assigns 4, 4, and 2\,req/s to Llama-3.1-8B, Qwen3-32B, and Llama-3.1-70B, respectively.
\sysname is evaluated over a range of request rates in the three datasets listed in Table~\ref{tab:slo_targets}, and the results are compared with the baselines.

Figure~\ref{fig:e2e_main} reports SLO attainment, request throughput, P90 TTFT, and P90 TPOT for \sysname and the baselines across the three datasets as the aggregate request rate increases.
\sysname sustains high SLO attainment across all experiments, with P90 TTFT and P90 TPOT remaining bounded under SLO as the rate increases. 
Averaged over all datasets and request rates, \sysname achieves 1.3$\times$, 4.1$\times$, and 1.5$\times$ higher SLO attainment over vanilla-vLLM, MuxServe$^\star$, and Prism$^\star$, respectively. 
Across all baselines, \sysname delivers an average 2.3${\times}$ improved SLO attainment and 1.2${\times}$ higher throughput. Section~\ref{subsec:ablation} further examines the contributions of contention-aware prediction and shared resource management.

The baselines degrade differently across datasets. For the long-context ArXiv and Azure-Code, Prism$^{\star}$ exhibits P90 TPOT above the SLO threshold while its P90 TTFT remains within the SLO, indicating that decode latency is the primary limitation in these workloads. Its chunked prefill co-batches prefill operations with ongoing decodes, increasing decode-step latency.
In contrast, ShareGPT features substantially shorter input lengths, making it less vulnerable to this issue at low loads. However, as the request rate escalates, a similar degradation pattern emerges: the P90 TPOT increases significantly due to the increased proportion of chunked prefill operations within the execution batches. 
MuxServe$^{\star}$ maintains comparatively low P90 TPOT through prefill-decode separation but exhibits elevated P90 TTFT across all datasets.
The greedy placement of MuxServe$^{\star}$ policy prioritizes the computation‑heavy prefill stage and executes prefills sequentially, leaving smaller models with degraded SLOs and pushing the tail of the TTFT distribution upward.

Figure~\ref{fig:req_stats} compares per-request TTFT and TPOT on Azure-Code at 5 and 15 req/s. At 15 req/s, the baselines show more requests exceeding the TTFT or TPOT thresholds, while \sysname maintains higher SLO attainment. 
For MuxServe$^\star$, in particular, the TTFT requests for Llama-3.1-8B and Llama-3.1-70B have TTFT exceeding the plotted limit of 16\,s. These results highlight the challenges of meeting SLOs when multiple serving instances share GPU resources. 
Although the baselines account for compute and memory capacity, they do not incorporate memory bandwidth contention into placement decisions. \sysname considers this additional factor when evaluating candidate placements, helping identify configurations that can sustain performance under co-location.

\begin{figure}[t]
    \centering
    \includegraphics[width=\columnwidth]{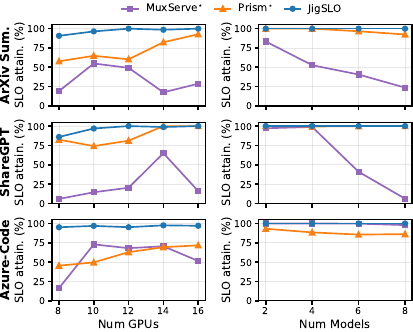}
    \caption{SLO attainments under GPU scaling (left) and model scaling (right) across three datasets. \sysname sustains high SLO attainment in all settings, while MuxServe$^\star$ and Prism$^\star$ degrade at smaller GPU counts and larger model counts.}
    \label{fig:e2e_scales}
\end{figure}

\smallskip
\noindent\textbf{GPU scaling.}
We evaluate each system's GPU efficiency by serving FP8-quantized variants of the three models (Llama-3.1-8B,  Qwen3-32B, Llama-3.1-70B) at fixed total request rates (the highest rate per dataset from the request rate scaling experiment). Each model's request rate remains fixed, while each system determines the number and configuration of its serving instances.
We vary the cluster size from 8 to 16 GPUs in increments of 2. This setup tests how many GPUs each system requires to sustain a given workload.


Figure~\ref{fig:e2e_scales} (left) shows SLO attainment versus cluster size. \sysname maintains high attainment across all datasets and cluster sizes (85.9\%–100\%), whereas MuxServe$^\star$ and Prism$^\star$ fall behind. Prism$^\star$'s KV-pressure-driven placement does not explicitly account for bandwidth contention, which can limit SLO attainment as fewer GPUs must accommodate the same workload.
MuxServe$^\star$'s SLO attainment does not improve monotonically with the number of GPUs: on ShareGPT it drops sharply at 16~GPUs. Its compute-greedy placement policy increases tensor parallelism to use the additional GPUs, changing the co-location configuration. This change may increase bandwidth contention among co-located instances, which the policy does not capture, contributing to the decline in SLO attainment.
In contrast, \sysname uses contention-aware latency estimates to exclude configurations predicted to violate SLOs and selects tensor-parallel configurations accordingly. On ArXiv Summarization and Azure-Code, \sysname matches the baselines' best SLO attainment at 16~GPUs using only 8~GPUs, reducing the GPU count by 50\% for the same workload.

\smallskip
\noindent\textbf{Model scaling.} We assess how each system copes with an increasing number of models by serving 2, 4, 6, and 8 Llama‑3.1‑8B (BF16) models on a 4‑GPU cluster. Each model receives an independent request stream at a per‑model rate of 0.8 req/s (ArXiv Summarization), 9 req/s (ShareGPT), and 3.5 req/s (Azure‑Code), so the total workload grows linearly with the model count. This configuration stresses the placement logic of each system and reveals how well it can preserve SLOs while selecting the placement candidate.

Figure~\ref{fig:e2e_scales} (right) reports SLO attainment as the number of co‑located models varies. MuxServe$^\star$ exhibits comparatively low SLO attainment. Its contention-agnostic placement greedily packs models onto fewer GPUs, which may increase bandwidth contention among co-located instances.
Prism$^\star$ drops from 100\% to 95\% (ArXiv Sum.) and 86.2\% (Azure‑Code). Its KV-pressure-driven placement leaves bandwidth contention unmodeled, which may contribute to this degradation.
In contrast, \sysname{}’s contention‑aware predictor estimates per‑instance latency for every candidate co‑location, enabling the placement algorithm to exclude configurations predicted to violate SLOs.
It achieves 100\% SLO attainment for every model count tested.

\subsection{Ablation Study}\label{subsec:ablation}
\begin{figure}[t]
    \centering
    \includegraphics[width=\columnwidth]{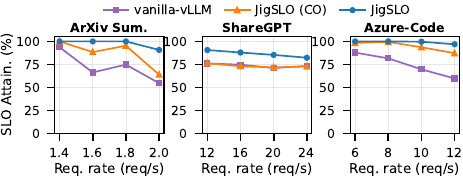}
    \caption{Ablation of \sysname{}’s \texttt{kvslab} and contention‑aware predictor across the three datasets. The transition vanilla-vLLM $\to$ \sysname (CO) isolates \texttt{kvslab}. \sysname (CO) $\to$ \sysname isolates the contention-aware predictor.}
    \label{fig:ablation}
\end{figure}
In this section, to evaluate each component's contribution, we compare three configurations that progressively enable shared KV cache management and contention-aware prediction:
vanilla-vLLM (CO placement, static KV partition), then \sysname (CO), which adds \texttt{kvslab}, then \sysname, which restores contention-aware placement. The CO variant uses the same placement algorithm with standalone latency estimates, retaining compute and memory-capacity constraints but omitting bandwidth contention. All three serve Qwen3-32B and Llama-3.1-8B on four GPUs at equal request rates, with instance counts determined by the placement produced for each configuration.



Figure~\ref{fig:ablation} reports SLO attainments for \sysname and the two ablation variants. Averaged over all datasets and request rates, \sysname (CO) achieves 1.16$\times$ the SLO attainment of vanilla-vLLM, while \sysname achieves 1.31$\times$ and 1.13$\times$ that of vanilla-vLLM and \sysname (CO), respectively.

\smallskip
\noindent\textbf{Effectiveness of shared KV cache management.} Our \texttt{kvslab} (vanilla-vLLM $\to$ \sysname (CO)) primarily benefits the long-context Arxiv and Azure-Code, where the SLO attainment gap widens with request rate.
This phenomenon occurs because these workloads feature long input lengths, which inherently lead to high KV cache footprints.
Under the static baseline (vanilla-vLLM), an instance can exhaust its assigned cache capacity while capacity reserved for other instances remains unused, delaying request processing. Our \texttt{kvslab} allows co-located instances to draw from a shared pool, improving cache utilization as their KV demands vary.

\smallskip
\noindent\textbf{Effectiveness of contention-aware prediction.} The contention-aware predictor (\sysname (CO) $\to$ \sysname) improves SLO attainment across all datasets, most prominently at high request rates.
Under heavy loads, CO's standalone latency estimates overlook bandwidth contention among co-located instances, overestimating the serving capacity.
In contrast, \sysname accounts for contention-induced latency increases when evaluating candidates, maintaining robust end-to-end performance across the evaluated workloads.


\begin{figure}[t]
    \centering
    \includegraphics[width=\columnwidth]{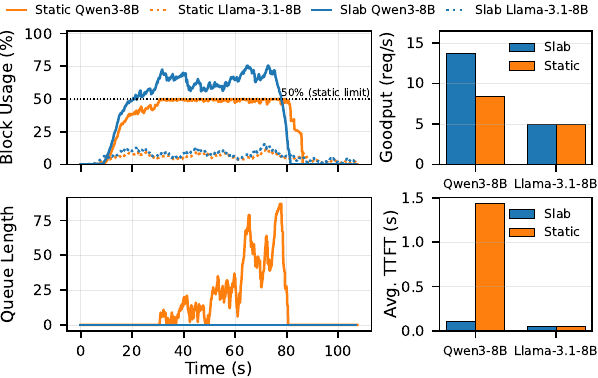}
    \caption{\texttt{kvslab} versus vLLM static KV partitioning deploying Qwen3-8B (15\,req/s) and Llama-3.1-8B (5\,req/s) on a single H100 GPU serving synthetic workload (input 1024, output 512). KV cache pool usage (Top-left) and queue length (bottom-left) timeline. Right: aggregate goodput and average TTFT per model.}
    \label{fig:kvslab_static}
\end{figure}

\subsection{Analysis of the KV Slab Manager}\label{sec:eval_kvslab}
In this subsection, we analyze the \texttt{kvslab} along two aspects: flexible shared KV cache coordination, and its overhead.

\smallskip
\noindent\textbf{Flexible shared KV cache management.}
We co-locate Qwen3-8B (15\,req/s) and Llama-3.1-8B (5\,req/s) on a single H100, using synthetic requests with 1024 input tokens and 512 output tokens. Figure~\ref{fig:kvslab_static} compares the \texttt{kvslab} against a static baseline that divides the total available KV cache capacity equally between two isolated per-instance pools.
We adopt the SLO targets from \S\ref{sec:setup} (TTFT/token$=$3\,ms, TPOT$=$50\,ms). Under static partitioning, Qwen3-8B saturates its 50\% quota and peaks at a $\sim$80-request backlog, inflating TTFT. Our \texttt{kvslab} instead pools the two instances' caches, letting Qwen3-8B reach 75\% usage with no backlog. This yields 1.6$\times$ higher goodput and 13.7$\times$ lower mean TTFT, with negligible impact on Llama-3.1-8B.

\begin{figure}[t]
    \centering
    \includegraphics[width=\columnwidth]{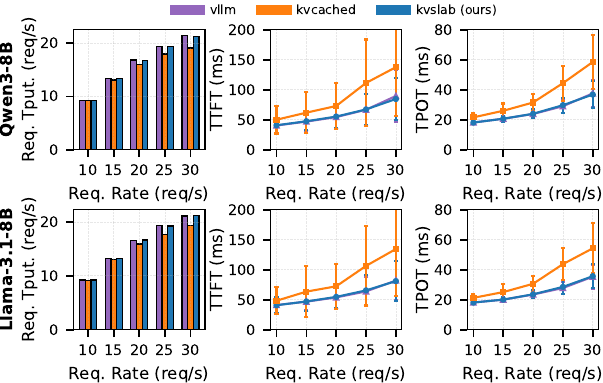}
    \caption{Comparison between our \texttt{kvslab} and \texttt{kvcached} deploying Qwen3-8B and Llama-3.1-8B on a single H100 GPU for serving ShareGPT dataset. \texttt{kvcached} exhibits higher TTFT and TPOT than \texttt{kvslab} as the request rate increases, while \texttt{kvslab} remains close to the vLLM baseline.}
    \vspace{-0.1in}
    \label{fig:kvslab_kvcached}
\end{figure}

\smallskip
\noindent\textbf{Serving overhead of KV cache management.} 
Figure~\ref{fig:kvslab_kvcached} compares our \texttt{kvslab} and \texttt{kvcached} against the vLLM baseline for co-located Qwen3-8B and Llama-3.1-8B on a single H100 serving ShareGPT. \texttt{kvcached} uses CUDA VMM to manage physical memory on demand. In our evaluated configuration, it incurs higher TTFT and TPOT than our \texttt{kvslab} as the request rate increases.
Our \texttt{kvslab} instead maps each instance's local block IDs onto physical blocks of the shared pool, avoiding driver calls. This adds only minimal coordination overhead, yielding negligible TTFT and TPOT increase over the vLLM baseline.

\subsection{Latency Prediction}\label{app:prediction_results}

\begin{table}[t]
    \centering
    \small
    \setlength{\tabcolsep}{2.5pt}
    \begin{tabular}{lcccccccc}
        \toprule
        & \multicolumn{4}{c}{\textbf{Standalone}}
        & \multicolumn{4}{c}{\textbf{Contended}} \\
        \cmidrule(lr){2-5} \cmidrule(lr){6-9}
        & \multicolumn{2}{c}{Prefill} & \multicolumn{2}{c}{Decode}
        & \multicolumn{2}{c}{Prefill} & \multicolumn{2}{c}{Decode} \\
        \cmidrule(lr){2-3} \cmidrule(lr){4-5} \cmidrule(lr){6-7} \cmidrule(lr){8-9}
        \textbf{Model} & BF16 & FP8 & BF16 & FP8 & BF16 & FP8 & BF16 & FP8 \\
        \midrule
        Llama-3.1-8B  & 8.43 & 9.31 & 5.40 & 6.53 & 14.79 & 14.46 & 12.56 & 12.45 \\
        Qwen3-32B     & 7.10 & 8.63 & 6.97 & 7.61 & 13.16 & 13.86 & 12.26 & 11.99 \\
        Llama-3.1-70B & 5.75 & 8.36 & 6.82 & 7.67 & 11.57 & 12.42 & 11.04 & 11.52 \\
        \bottomrule
    \end{tabular}
    \caption{Latency prediction MAPE (\%) for standalone and contended execution, across precisions and phases.}
    \vspace{-0.2in}
    \label{tab:prediction_mape}
\end{table}


We evaluate the predictor (\S\ref{sec:latency_prediction}) on three models in two precisions for both prefill and decode phases, sweeping SM allocation, input length, and batch size. For standalone latency, we report the regression fitting error against the standalone measurements used to fit the model. For contended latency, we compare analytical predictions against measured co-located execution latencies, which are not used for fitting. Each contended entry aggregates errors across the evaluated co-location configurations involving the target.
Table~\ref{tab:prediction_mape} reports the mean absolute percentage error (MAPE). Standalone fitting error ranges from 5.40\% to 9.31\%, while contended prediction error ranges from 11.04\% to 14.79\%. These results show that the analytical contention model predicts co-located latency across the evaluated configurations without fitting to co-located measurements.

Prior work drives configuration search with similar latency models fitted to standalone profiles~\cite{agrawal2024vidur, zhong2024distserve}. Those models assume the instance owns its GPU and account for neither partial SM allocation nor bandwidth contention among co-located instances, so the contended entries in Table~\ref{tab:prediction_mape} have no prior counterpart. The end-to-end results in~\S\ref{subsec:end-to-end_performance} and~\S\ref{subsec:ablation} indicate that this accuracy is sufficient for placement, as the co-locations selected under these predictions sustain high SLO attainment.


\section{Related work}

\smallskip
\noindent\textbf{Multi-LLM spatial sharing.}
MuxServe~\citep{duan2024muxserve} multiplexes LLMs spatially and temporally but places them based on profiled performance rather than an explicit contention model, and Prism~\citep{yu2025prism} balances per-GPU KV memory pressure against SLO headroom, modeling capacity rather than bandwidth. Earlier multi-model serving work targets placement and replication~\citep{li2023alpaserve, gujarati2020clockwork, zhang2023shepherd}, profiled interference~\citep{choi2022stsharing}, cold-start locality~\citep{fu2024serverlessllm}, and shared-backbone adapter multiplexing~\citep{chen2024punica, sheng2024slora, wu2024dlora}. KV cache managers reduce fragmentation within one instance via paging~\citep{kwon2023vllm}, virtual memory~\citep{prabhu2025vattention, xu2024vtensor}, disaggregated pools~\citep{qin2025mooncake}, or heterogeneity-aware allocation~\citep{zhang2025jenga}, but assume the instance owns its GPU and cannot reclaim memory across co-located models.

\smallskip
\noindent\textbf{Single-model prefill-decode multiplexing.} A parallel line co-locates one model's prefill and decode on a GPU: Bullet~\citep{lin2026bullet} predicts kernel latency under partial SM allocation with a per-configuration linear calibration, and MuxWise~\citep{chen2026muxwise} reserves best-fit SMs to protect decode SLOs. Calibration is tractable there because a model's prefill is paired with its decode, leaving only workload size and SM allocation to sweep. Across heterogeneous models, each instance in a co-location independently ranges over (model, phase, precision, workload size, SM allocation), and each new model must be measured against every existing configuration, so per-configuration calibration becomes infeasible. \sysname instead requires no measurement beyond standalone profiles, composing per-instance memory-utilization predictions analytically.

\smallskip
\noindent\textbf{GPU contention estimation.} Orion~\citep{eurosys2024orion} and Reef~\citep{osdi2022reef} mitigate interference by scheduling kernel execution timing, and Mudi~\citep{eurosys2025mudi} estimates inference--training interference with a piecewise-linear model. These systems split workloads into high-priority and best-effort tiers. \sysname instead co-locates instances that must all meet their SLOs, and predicts contention rather than reacting to it at runtime.

\section{Conclusion}\label{sec:conclusion}

We presented \sysname, a contention-aware multi-LLM serving system that makes GPU spatial sharing reliable under SLOs through three contributions: a latency predictor that reduces bandwidth contention among co-located instances to a single mechanism (time-shared memory access), an SLO-aware placement algorithm that uses these predictions to co-locate instances with complementary memory-access characteristics, and a shared resource manager that eliminates weight duplication and avoids KV cache fragmentation. Across three datasets and three state-of-the-art baselines, \sysname achieves 2.3$\times$ higher SLO attainment and 1.2$\times$ higher throughput on average, showing that contention-aware placement and shared resource management together make co-located multi-LLM serving cost-efficient.




\bibliographystyle{ACM-Reference-Format}
\bibliography{reference}

@article{openai2024o1,
  title={OpenAI o1 System Card},
  author={OpenAI},
  year={2024},
  note={\url{https://openai.com/index/openai-o1-system-card/}}
}

@article{deepseek2025r1,
  title={DeepSeek-R1: Incentivizing Reasoning Capability in LLMs via Reinforcement Learning},
  author={DeepSeek-AI},
  year={2025},
  journal={arXiv preprint arXiv:2501.12948}
}

@misc{claudecode,
  title={Claude Code},
  author={Anthropic},
  year={2025},
  note={\url{https://www.anthropic.com/claude-code}}
}

@misc{openai2025codex,
  title={Introducing Codex},
  author={OpenAI},
  year={2025},
  note={\url{https://openai.com/index/introducing-codex/}}
}

@inproceedings{kwon2023vllm,
  title={Efficient Memory Management for Large Language Model Serving with PagedAttention},
  author={Woosuk Kwon and Zhuohan Li and Siyuan Zhuang and Ying Sheng and Lianmin Zheng and Cody Hao Yu and Joseph E. Gonzalez and Hao Zhang and Ion Stoica},
  booktitle={Proceedings of the ACM SIGOPS 29th Symposium on Operating Systems Principles},
  year={2023}
}

@article{zheng2023sglang,
  title={Efficiently Programming Large Language Models using SGLang.},
  author={Zheng, Lianmin and Yin, Liangsheng and Xie, Zhiqiang and Huang, Jeff and Sun, Chuyue and Yu, Cody Hao and Cao, Shiyi and Kozyrakis, Christos and Stoica, Ion and Gonzalez, Joseph E and others},
  year={2023},
  publisher={arXiv}
}

@inproceedings{duan2024muxserve,
  title={MuxServe: flexible spatial-temporal multiplexing for multiple LLM serving},
  author={Duan, Jiangfei and Lu, Runyu and Duanmu, Haojie and Li, Xiuhong and Zhang, Xingcheng and Lin, Dahua and Stoica, Ion and Zhang, Hao},
  booktitle={Proceedings of the 41st International Conference on Machine Learning},
  pages={11905--11917},
  year={2024}
}

@inproceedings {yu2025prism,
author = {Shan Yu and Yifan Qiao and Mingyuan Ma and Yangmin Li and Shuo Yang and Xinyuan Tong and Yang Wang and Zhiqiang Xie and Yuwei An and Shiyi Cao and Ke Bao and Deepak Vij and Xiaoning Ding and Yichen Wang and Qingda Lu and Zhong Wang and Gao Gao and Harry Xu and Junyi Shu and Jiarong Xing and Ying Sheng},
title = {Prism: {Cost-Efficient} {Multi-LLM} Serving via {GPU} Memory Ballooning},
booktitle = {20th USENIX Symposium on Operating Systems Design and Implementation (OSDI 26)},
year = {2026},
isbn = {978-1-939133-55-7},
pages = {75--93},
url = {https://www.usenix.org/conference/osdi26/presentation/yu-shan},
publisher = {USENIX Association},
month = jul
}

@inproceedings{lin2026bullet,
  title={Bullet: Boosting gpu utilization for llm serving via dynamic spatial-temporal orchestration},
  author={Lin, Zejia and Xu, Hongxin and Chen, Guanyi and Chen, Zhiguang and Lu, Yutong and Zhang, Xianwei},
  booktitle={Proceedings of the 31st ACM International Conference on Architectural Support for Programming Languages and Operating Systems (ASPLOS 26), Volume 2},
  pages={290--306},
  year={2026}
}

@inproceedings{agrawal2024sarathi,
  title={Taming $\{$Throughput-Latency$\}$ tradeoff in $\{$LLM$\}$ inference with $\{$Sarathi-Serve$\}$},
  author={Agrawal, Amey and Kedia, Nitin and Panwar, Ashish and Mohan, Jayashree and Kwatra, Nipun and Gulavani, Bhargav and Tumanov, Alexey and Ramjee, Ramachandran},
  booktitle={18th USENIX symposium on operating systems design and implementation (OSDI 24)},
  pages={117--134},
  year={2024}
}

@inproceedings{patel2024splitwise,
  title={Splitwise: Efficient generative llm inference using phase splitting},
  author={Patel, Pratyush and Choukse, Esha and Zhang, Chaojie and Shah, Aashaka and Goiri, {\'I}{\~n}igo and Maleki, Saeed and Bianchini, Ricardo},
  booktitle={2024 ACM/IEEE 51st Annual International Symposium on Computer Architecture (ISCA 24)},
  pages={118--132},
  year={2024},
  organization={IEEE}
}

@inproceedings{zhong2024distserve,
  title={DistServe: Disaggregating prefill and decoding for goodput-optimized large language model serving},
  author={Zhong, Yinmin and Liu, Shengyu and Chen, Junda and Hu, Jianbo and Zhu, Yibo and Liu, Xuanzhe and Jin, Xin and Zhang, Hao},
  booktitle={18th USENIX Symposium on Operating Systems Design and Implementation (OSDI 24)},
  pages={193--210},
  year={2024}
}

@inproceedings{chen2026muxwise,
  title={Towards High-Goodput LLM Serving with Prefill-decode Multiplexing},
  author={Chen, Yukang and Cui, Weihao and Zhao, Han and Xu, Ziyi and Fan, Xiaoze and Chen, Xusheng and Zhou, Yangjie and Sun, Shixuan and He, Bingsheng and Chen, Quan},
  booktitle={Proceedings of the 31st ACM International Conference on Architectural Support for Programming Languages and Operating Systems (ASPLOS 26), Volume 2},
  pages={2030--2047},
  year={2026}
}

@inproceedings{xiang2025aegaeon,
  title={Aegaeon: Effective GPU pooling for concurrent LLM serving on the market},
  author={Xiang, Yuxing and Li, Xue and Qian, Kun and Yang, Yufan and Zhu, Diwen and Yu, Wenyuan and Zhai, Ennan and Liu, Xuanzhe and Jin, Xin and Zhou, Jingren},
  booktitle={Proceedings of the ACM SIGOPS 31st Symposium on Operating Systems Principles (SOSP 25)},
  pages={1030--1045},
  year={2025}
}

@inproceedings{elvinger2025understanding,
  title={Understanding GPU Resource Interference One Level Deeper},
  author={Elvinger, Paul and Strati, Foteini and Jerger, Natalie Enright and Klimovic, Ana},
  booktitle={Proceedings of the 2025 ACM Symposium on Cloud Computing},
  pages={687--694},
  year={2025}
}

@misc{yuan2024llm,
      title={LLM Inference Unveiled: Survey and Roofline Model Insights}, 
      author={Zhihang Yuan and Yuzhang Shang and Yang Zhou and Zhen Dong and Chenhao Xue and Bingzhe Wu and Zhikai Li and Qingyi Gu and Yong Jae Lee and Yan Yan and Beidi Chen and Guangyu Sun and Kurt Keutzer},
      year={2024},
      eprint={2402.16363},
      archivePrefix={arXiv},
      primaryClass={cs.CL}
}

@article{agrawal2024vidur,
  title={Vidur: A large-scale simulation framework for llm inference},
  author={Agrawal, Amey and Kedia, Nitin and Mohan, Jayashree and Panwar, Ashish and Kwatra, Nipun and Gulavani, Bhargav S and Ramjee, Ramachandran and Tumanov, Alexey},
  journal={Proceedings of Machine Learning and Systems},
  volume={6},
  pages={351--366},
  year={2024}
}

@inproceedings{qin2025mooncake,
  author    = {Ruoyu Qin and Zheming Li and Weiran He and Jialei Cui and Feng Ren and Mingxing Zhang and Yongwei Wu and Weimin Zheng and Xinran Xu},
  title     = {Mooncake: Trading More Storage for Less Computation {\textemdash} A {KVCache-centric} Architecture for Serving {LLM} Chatbot},
  booktitle = {23rd USENIX Conference on File and Storage Technologies (FAST 25)},
  year      = {2025},
  isbn      = {978-1-939133-45-8},
  address   = {Santa Clara, CA},
  pages     = {155--170},
  url       = {https://www.usenix.org/conference/fast25/presentation/qin},
  publisher = {USENIX Association},
  month     = {feb},
}

@inproceedings{eurosys2024orion,
author = {Strati, Foteini and Ma, Xianzhe and Klimovic, Ana},
title = {Orion: Interference-aware, Fine-grained GPU Sharing for ML Applications},
year = {2024},
isbn = {9798400704376},
publisher = {Association for Computing Machinery},
address = {New York, NY, USA},
url = {https://doi.org/10.1145/3627703.3629578},
doi = {10.1145/3627703.3629578},
booktitle = {Proceedings of the Nineteenth European Conference on Computer Systems},
pages = {1075–1092},
numpages = {18},
location = {Athens, Greece},
series = {EuroSys '24}
}

@inproceedings {osdi2022reef,
author = {Mingcong Han and Hanze Zhang and Rong Chen and Haibo Chen},
title = {Microsecond-scale Preemption for Concurrent {GPU-accelerated} {DNN} Inferences},
booktitle = {16th USENIX Symposium on Operating Systems Design and Implementation (OSDI 22)},
year = {2022},
isbn = {978-1-939133-28-1},
address = {Carlsbad, CA},
pages = {539--558},
url = {https://www.usenix.org/conference/osdi22/presentation/han},
publisher = {USENIX Association},
month = jul
}

@inproceedings{eurosys2025mudi,
author = {Chen, Wenyan and Lu, Chengzhi and Xu, Huanle and Ye, Kejiang and Xu, Chengzhong},
title = {Multiplexing Dynamic Deep Learning Workloads with SLO-awareness in GPU Clusters},
year = {2025},
isbn = {9798400711961},
publisher = {Association for Computing Machinery},
address = {New York, NY, USA},
url = {https://doi.org/10.1145/3689031.3696074},
doi = {10.1145/3689031.3696074},
booktitle = {Proceedings of the Twentieth European Conference on Computer Systems},
pages = {589–604},
numpages = {16},
location = {Rotterdam, Netherlands},
series = {EuroSys '25}
}

@article{xing2025towards,
  title={Towards Efficient and Practical GPU Multitasking in the Era of LLM},
  author={Xing, Jiarong and Qiao, Yifan and Mo, Simon and Cui, Xingqi and Sela, Gur-Eyal and Zhou, Yang and Gonzalez, Joseph and Stoica, Ion},
  journal={arXiv preprint arXiv:2508.08448},
  year={2025}
}

@article{tan2021serving,
  title={Serving dnn models with multi-instance gpus: A case of the reconfigurable machine scheduling problem},
  author={Tan, Cheng and Li, Zhichao and Zhang, Jian and Cao, Yu and Qi, Sikai and Liu, Zherui and Zhu, Yibo and Guo, Chuanxiong},
  journal={arXiv preprint arXiv:2109.11067},
  year={2021}
}

@inproceedings{lee2024parvagpu,
  title={ParvaGPU: Efficient spatial GPU sharing for large-scale DNN inference in cloud environments},
  author={Lee, Munkyu and Seong, Sihoon and Kang, Minki and Lee, Jihyuk and Na, Gap-Joo and Chun, In-Geol and Nikolopoulos, Dimitrios and Hong, Cheol-Ho},
  booktitle={SC24: International Conference for High Performance Computing, Networking, Storage and Analysis},
  pages={1--14},
  year={2024},
  organization={IEEE}
}

@misc{nvidia_mps,
author = "{NVIDIA}",
title = "{Multi-Process Service}",
howpublished = "\url{https://docs.nvidia.com/deploy/mps/index.html}",
year = 2026,
}

@misc{nvidia_mig,
author = "{NVIDIA}",
title = "{Multi-Instance GPU}",
howpublished = "\url{https://www.nvidia.com/en-us/technologies/multi-instance-gpu/}",
year = 2026,
}

@misc{sharegpt2023,
    author = {{ShareGPT Team}},
    title = {{ShareGPT}},
    year = {2023},
    howpublished = {\url{https://sharegpt.com/}},
}

@inproceedings{cohan2018arxivsum,
  title     = {A Discourse-Aware Attention Model for Abstractive 
               Summarization of Long Documents},
  author    = {Cohan, Arman and Dernoncourt, Franck and Kim, Doo Soon and 
               Bui, Trung and Kim, Seokhwan and Chang, Walter and 
               Goharian, Nazli},
  booktitle = {Proceedings of the 2018 Conference of the North {A}merican 
               Chapter of the Association for Computational Linguistics: 
               Human Language Technologies, Volume 2 (Short Papers)},
  pages     = {615--621},
  year      = {2018},
  publisher = {Association for Computational Linguistics},
  url       = {https://aclanthology.org/N18-2097},
  doi       = {10.18653/v1/N18-2097},
}

@misc{VMM,
author = "{NVIDIA}",
title = "{CUDA Toolkit Documentation: Virtual Memory Management}",
howpublished = "\url{https://docs.nvidia.com/cuda/cuda-driver-api/group__CUDA__VA.html}",
year = 2026,
}

@misc{ZMQ,
author = "{ZeroMQ}",
title = "{ZeroMQ Website}",
howpublished = "\url{https://zeromq.org/}",
year = 2026,
}

@inproceedings{li2023alpaserve,
  author    = {Zhuohan Li and Lianmin Zheng and Yinmin Zhong and Vincent Liu and Ying Sheng and Xin Jin and Yanping Huang and Zhifeng Chen and Hao Zhang and Joseph E. Gonzalez and Ion Stoica},
  title     = {{AlpaServe}: Statistical Multiplexing with Model Parallelism for Deep Learning Serving},
  booktitle = {17th USENIX Symposium on Operating Systems Design and Implementation (OSDI 23)},
  year      = {2023},
  isbn      = {978-1-939133-34-2},
  address   = {Boston, MA},
  pages     = {663--679},
  url       = {https://www.usenix.org/conference/osdi23/presentation/li-zhuohan},
  publisher = {USENIX Association},
  month     = jul
}

@inproceedings{fu2024serverlessllm,
  author    = {Yao Fu and Leyang Xue and Yeqi Huang and Andrei-Octavian Brabete and Dmitrii Ustiugov and Yuvraj Patel and Luo Mai},
  title     = {{ServerlessLLM}: {Low-Latency} Serverless Inference for Large Language Models},
  booktitle = {18th USENIX Symposium on Operating Systems Design and Implementation (OSDI 24)},
  year      = {2024},
  isbn      = {978-1-939133-40-3},
  address   = {Santa Clara, CA},
  pages     = {135--153},
  url       = {https://www.usenix.org/conference/osdi24/presentation/fu},
  publisher = {USENIX Association},
  month     = jul
}

@inproceedings{wu2024dlora,
  author    = {Bingyang Wu and Ruidong Zhu and Zili Zhang and Peng Sun and Xuanzhe Liu and Xin Jin},
  title     = {{dLoRA}: Dynamically Orchestrating Requests and Adapters for {LoRA} {LLM} Serving},
  booktitle = {18th USENIX Symposium on Operating Systems Design and Implementation (OSDI 24)},
  year      = {2024},
  isbn      = {978-1-939133-40-3},
  address   = {Santa Clara, CA},
  pages     = {911--927},
  url       = {https://www.usenix.org/conference/osdi24/presentation/wu-bingyang},
  publisher = {USENIX Association},
  month     = jul
}

@inproceedings{sheng2024slora,
  author    = {Ying Sheng and Shiyi Cao and Dacheng Li and Coleman Hooper and Nicholas Lee and Shuo Yang and Christopher Chou and Banghua Zhu and Lianmin Zheng and Kurt Keutzer and Joseph E. Gonzalez and Ion Stoica},
  title     = {{SLoRA}: Scalable Serving of Thousands of {LoRA} Adapters},
  booktitle = {Proceedings of the 7th Annual Conference on Machine Learning and Systems (MLSys 24)},
  year      = {2024},
  address   = {Santa Clara, CA}
}

@inproceedings{chen2024punica,
  author    = {Lequn Chen and Zihao Ye and Yongji Wu and Danyang Zhuo and Luis Ceze and Arvind Krishnamurthy},
  title     = {Punica: Multi-Tenant {LoRA} Serving},
  booktitle = {Proceedings of the 7th Annual Conference on Machine Learning and Systems (MLSys 24)},
  year      = {2024},
  address   = {Santa Clara, CA}
}

@inproceedings{gujarati2020clockwork,
  author    = {Arpan Gujarati and Reza Karimi and Safya Alzayat and Wei Hao and Antoine Kaufmann and Ymir Vigfusson and Jonathan Mace},
  title     = {Serving {DNNs} like Clockwork: Performance Predictability from the Bottom Up},
  booktitle = {14th USENIX Symposium on Operating Systems Design and Implementation (OSDI 20)},
  year      = {2020},
  isbn      = {978-1-939133-19-9},
  pages     = {443--462},
  url       = {https://www.usenix.org/conference/osdi20/presentation/gujarati},
  publisher = {USENIX Association},
  month     = nov
}

@inproceedings{choi2022stsharing,
  author    = {Seungbeom Choi and Sunho Lee and Yeonjae Kim and Jongse Park and Youngjin Kwon and Jaehyuk Huh},
  title     = {Serving Heterogeneous Machine Learning Models on {Multi-GPU} Servers with {Spatio-Temporal} Sharing},
  booktitle = {2022 USENIX Annual Technical Conference (USENIX ATC 22)},
  year      = {2022},
  address   = {Carlsbad, CA},
  pages     = {199--216},
  url       = {https://www.usenix.org/conference/atc22/presentation/choi-seungbeom},
  publisher = {USENIX Association},
  month     = jul
}

@inproceedings{zhang2023shepherd,
  author    = {Hong Zhang and Yupeng Tang and Anurag Khandelwal and Ion Stoica},
  title     = {{SHEPHERD}: Serving {DNNs} in the Wild},
  booktitle = {20th USENIX Symposium on Networked Systems Design and Implementation (NSDI 23)},
  year      = {2023},
  isbn      = {978-1-939133-33-5},
  address   = {Boston, MA},
  pages     = {787--808},
  url       = {https://www.usenix.org/conference/nsdi23/presentation/zhang-hong},
  publisher = {USENIX Association},
  month     = apr
}

@inproceedings{prabhu2025vattention,
  author    = {Ramya Prabhu and Ajay Nayak and Jayashree Mohan and Ramachandran Ramjee and Ashish Panwar},
  title     = {{vAttention}: Dynamic Memory Management for Serving {LLMs} without {PagedAttention}},
  booktitle = {Proceedings of the 30th ACM International Conference on Architectural Support for Programming Languages and Operating Systems, Volume 1 (ASPLOS 25)},
  year      = {2025},
  publisher = {Association for Computing Machinery},
  address   = {New York, NY, USA},
  location  = {Rotterdam, The Netherlands},
  doi       = {10.1145/3669940.3707256}
}

@inproceedings{zhang2025jenga,
  author    = {Chen Zhang and Kuntai Du and Shu Liu and Woosuk Kwon and Xiangxi Mo and Yufeng Wang and Xiaoxuan Liu and Kaichao You and Zhuohan Li and Mingsheng Long and Jidong Zhai and Joseph E. Gonzalez and Ion Stoica},
  title     = {Jenga: Effective Memory Management for Serving {LLM} with Heterogeneity},
  booktitle = {Proceedings of the ACM SIGOPS 31st Symposium on Operating Systems Principles (SOSP 25)},
  year      = {2025},
  publisher = {Association for Computing Machinery},
  address   = {New York, NY, USA},
  doi       = {10.1145/3731569.3764823}
}

@article{xu2024vtensor,
  author  = {Jiale Xu and Rui Zhang and Cong Guo and Weiming Hu and Zihan Liu and Feiyang Wu and Yu Feng and Shixuan Sun and Changxu Shao and Yuhong Guo and Junping Zhao and Ke Zhang and Minyi Guo and Jingwen Leng},
  title   = {{vTensor}: Flexible Virtual Tensor Management for Efficient {LLM} Serving},
  journal = {arXiv preprint arXiv:2407.15309},
  year    = {2024}
}



\end{document}